\documentclass[preprint, 12pt]{elsarticle}

\usepackage{amssymb}
\usepackage{amsmath}
\usepackage{xcolor}
\usepackage{breqn}
\usepackage{amsmath}
\usepackage[symbol]{footmisc}
\usepackage{comment}

\begin{document}

\begin{frontmatter}



\title{Dynamics of a dark soliton in a curved 1D Bose-Einstein condensate }


\author[1]{Jorge A. G. Attie}
\ead{jorge.attie@usp.br}
\author[1]{Emanuel A. L. Henn\corref{cor1}}
\ead{ehenn@ifsc.usp.br}
\cortext[cor1]{Corresponding author}
\affiliation[1]{organization={São Carlos Institute of Physics, University of São Paulo, IFSC-USP},
            postcode={13566-590}, 
            city={São Carlos},
            state={São Paulo},
            country={Brazil}}

\begin{abstract}
We investigate the nonlinear dynamics of dark solitons in a one-dimensional Bose-Einstein condensate confined to a curved geometry. Using the Gross-Pitaevskii equation in curvilinear coordinates and a perturbative expansion in the local curvature, we derive a set of coupled evolution equations for the soliton velocity and the curvature. For the case of constant curvature, such as circular geometries, the soliton dynamics is governed solely by the initial velocity and curvature. Remarkably, the soliton travels a nearly constant angular trajectory across two orders of magnitude in curvature, suggesting an emergent conserved quantity, independent of its initial velocity. We extend our analysis to elliptical trajectories with spatially varying curvature and show that soliton dynamics remain determined by the local curvature profile. In these cases, the model of effective constant curvature describes accurately the dynamics given the local curvature has smooth variation. When the soliton crosses regions of rapid curvature variation and/or non-monotonic behavior, the model fails to describe to soliton dynamics, although the overall behavior can still be fully mapped to the curvature profile. Our results provide a quantitative framework for understanding the role of geometry in soliton dynamics and pave the way for future studies of nonlinear excitations in curved quantum systems.\end{abstract}




\end{frontmatter}






\section{Introduction}
\label{Introduction}

Nonlinear systems of various origins are generally known to support the existence of solitons. Solitonic configurations play a significant role in a wide range of physical systems, including nuclear matter \cite{PhysRevD.46.3903, Ohno}, electromagnetic fields \cite{agrawal}, gravitational systems \cite{Mielke, Kunz}, classical and quantum fluids \cite{fluids} and even biological systems \cite{bio}.

In recent years, increasing attention has been devoted to soliton solutions in the context of bosonic quantum fluids \cite{Frantzeskakis2010}. In particular, Bose-Einstein condensates (BECs), which are described by the Gross-Pitaevskii equation (GPE) \cite{Pitaevskii2016}, support the formation of self-localized states, manifesting either as a density peak or a density dip, that propagate with constant shape in a homogeneous background \cite{Busch2001}. The former corresponds to a bright soliton \cite{Strecker2002}, while the latter represents a grey or dark soliton \cite{Burger1999}. Since the early days following the experimental realization of BECs \cite{Anderson1995}, solitons have remained a central topic in both theoretical and experimental studies \cite{Dalfovo1999,Kivshar2014}. This growing interest is partly due to the remarkable tunability of the nonlinearity in the GPE, which arises from mean-field two-body interactions and can be controlled experimentally \cite{Cornell2002}. 

Interestingly, although stemming from a completely different physical context, the propagation of light in nonlinear optical media is often described by the same mathematical equation as bosonic quantum fluids: the nonlinear Schrödinger equation \cite{Kivshar1989}. This formal analogy enables a unified approach to the physics of solitons across different systems.

Moreover, light propagating through a medium can serve as an analogue for a field perturbation traveling through curved spacetime, akin to a gravitational background. Batz and Peschel \cite{WOS:000278140000167} explored optical solitons in media with constant curvature, showing that the curvature of space fundamentally influences the existence and stability of solitons. Specifically, they found that positive curvature supports stable solitons with intricate internal structures, while negative curvature leads to the decay of all localized structures, irrespective of the nonlinearity strength. Further expanding on this, Spengler et al.\cite{WOS:001012312400001} investigated the behavior of optical solitons in nonlinear media subjected to curved spacetime, such as near the Earth’s surface under gravitational influence. Notably, they demonstrated that light propagation in curved spacetime can be equivalently described as propagation in an inhomogeneous effective medium in flat spacetime.

Until recently, the effects of curvature represented a striking asymmetry between the treatment of Bose-Einstein condensates (BECs) and optical fields, as investigations of BECs and their excitations on curved manifolds were largely absent from the literature \cite{Fetter2003}. This scenario has changed in recent years, driven by growing interest in the behavior of atomic superfluids in non-flat geometries, where systems are characterized by local curvature, periodic boundary conditions, and non-trivial topologies. Examples include toroidal and ring-shaped traps \cite{Ryu2007,Eckel2014} and curved waveguides \cite{Eisenberg1998,Morandotti1999} in one dimension, as well as bubble-like traps in thin and thick shells in two and three dimensions \cite{padavic2020}.

This surge in interest has led to investigations into how curvature and topology influence key properties of quantum gases, including the nature of the Bose-Einstein condensation transition and the superfluid phase \cite{tononi2023}. Researchers have explored how collective excitations are modified in curved geometries and how their spectral signatures may serve as indicators of the topological transition, such as the evolution from filled to hollow BECs \cite{sun_a, sun_b}. Additionally, vortex dynamics are also significantly affected by curvature, topology and dimensionality \cite{tononi2023,padavic2020}. Additional non-linear contributions like dipole-dipole mean field interaction potential further modify the ground state and collective excitations in the presence of curvature in thin shell potentials \cite{Diniz:2020aa}.

This line of research has been fueled by experimental advances that enable the realization and control of curved trapping potentials \cite{perrin2021, rey2022, yanliang2022}. Notably, ring traps and atomic waveguides \cite{amico2021} have been successfully implemented, and recent breakthroughs have enabled the creation of tunable shell-like potentials \cite{dubessy2025}. These include experiments conducted aboard the NASA Cold Atom Laboratory (CAL) \cite{lundblad} on the International Space Station, where microgravity conditions allow the formation of large superfluid bubbles with variable dimensions. Similar shell-like structures have also been realized in terrestrial laboratories using immiscible mixtures of ultracold atoms and finely tuned trapping configurations. Comprehensive reviews published recently offer a detailed overview of this rapidly evolving field \cite{dubessy2025}.

As for the study of solitons in Bose-Einstein condensates (BECs) under the influence of curvature, to the best of our knowledge, the only prior treatment was carried out by Mironov and Smirnov  \cite{mironov2010dynamics,PhysRevA.85.053620}. Their work employed an asymptotic approach to analyze the dynamics of a dark soliton propagating along a curved line with arbitrary but small curvature. In this framework, the dark soliton model was used to reveal the emergence of vortex filaments in quasi-two-dimensional Bose gases, and to explore the role such vortices play in the transition from a coherent BEC to a turbulent superfluid state.

In the present work, we build upon the same formalism introduced in Ref.\cite{mironov2010dynamics}, with a particular focus on the dynamics of solitons along curved trajectories with smooth but well-defined curvature, such as closed circles and ellipses. This corresponds to studying the dynamics of solitonic excitations in a bosonic superfluid confined to a one-dimensional ring-shaped trap or propagating along a path on the surface of a curved, closed geometry, such as a thin-shell BEC.

The paper is organized as follows: we begin with a brief review of the theoretical framework developed in Ref.\cite{mironov2010dynamics} and establish the fundamental equations of motion for a dark soliton propagating along a one-dimensional line with arbitrary curvature. We then examine the soliton dynamics under conditions of constant curvature, such as motion along a circular trajectory. For curvatures spanning several orders of magnitude, we demonstrate the emergence of a universal behavior across a wide range of constant curvature values. Moreover, in this specified range, we show that the soliton travels a constant angular displacement, regardless of the curvature or initial velocity. Subsequently, we analyze the motion of a dark soliton along an elliptical profile, where the local curvature varies with position. We show that the local curvature fully determines the soliton dynamics. Indeed, depending upon the curvature profile along the soliton trajectory, the soliton behavior can be described by an effective constant curvature. 
We summarize our main findings at the final section.

\section{Model: solitons in BECs with curvature}


In the absence of an external potential, the dynamics of the condensate wavefunction $\Psi(\vec{r},t)$ is governed by the Gross-Pitaevskii equation (GPE) in its dimensional form: 

\begin{equation}
i\hbar\frac{\partial\Psi}{\partial t} = -\frac{\hbar^{2}}{2m}\nabla^{2}\Psi + g|\Psi|^{2}\Psi,
\label{GPE_dimensional}
\end{equation}
where $m$ is the atomic mass and $g$ is the effective interaction strength. The GPE can be conveniently expressed in its adimensional representation like (see \ref{app:admi}).

\begin{equation}
    i\frac{\partial\Psi}{\partial t} + \frac{1}{2}\nabla^{2}\Psi + |\Psi|^{2}\Psi = 0.
\end{equation}

We are interested in the dynamics of a dark soliton representing a localized density dip superimposed on a uniform background. Far from the soliton core, the density approaches the background value, such that $|\Psi|^{2}=1$. Consequently, the GPE can be conveniently rewritten to explicitly account for deviations from this background as:

\begin{equation}
    i\frac{\partial\Psi}{\partial t} + \frac{1}{2}\nabla^{2}\Psi + (1-|\Psi|^{2})\Psi = 0.
    \label{GPE_dimensionless}
\end{equation}

This equation admits static solutions in the form of dark solitons, whose wavefunction is typically written as:

\begin{equation}
\Psi_{0}(x,t) = \frac{1}{\Lambda}\tanh\left(\frac{x - vt}{\Lambda}\right) + iv,
\label{DS_wavefunction_0}
\end{equation}
where $v$ is the velocity of the dark soliton and $\Lambda = (1-v^{2})^{-1/2}$ represents its width. However, our interest lies in more complex structures, the so-called \textit{curved dark solitons}, which although locally resembling the planar GPE solution, are confined along the smooth curve $\vec{r}(s,t)$, where $s$ in the arc length.

For the dark soliton approximation to remain valid, the width $\Lambda$ must vary smoothly along the curve, and the local radius of curvature must remain large. The latter can be expressed as $R(s,t) = 1/k(s,t)$, where $k(s,t)$ is the curvature, which is described in polar coordinates as follow:

\begin{equation}
    k\left(\rho,\phi\right) = \frac{\rho^{2} + \left(\partial_{\phi\phi}\rho\right)^{2} - \rho\partial^{2}_{\phi\phi}\rho}{\left[\rho^{2}+\left(\partial^{2}_{\phi\phi}\rho\right)^{2}\right]^{3/2}},
    \label{curvature_polar_coordinates}
\end{equation}
with $\left(\rho,\phi\right)$ being the radial and angular polar coordinates, respectively. The aforementioned condition are satisfied if the following inequalities hold:

\begin{equation}
    |\partial_{s}\Lambda(s,t)| << 1 
    \label{smoothly_conditions}
\end{equation}
and
\begin{equation}
    |k(s,t)\Lambda(s,t)| << 1.
\end{equation}
The GPE in curvilinear coordinates can be written as (see \ref{app:GPEcurve}):
\begin{equation}
i\frac{\partial \Psi}{\partial t} + \frac{1}{2}\frac{\partial^{2}\Psi}{\partial s^{2}} + \frac{k}{2}\frac{\partial^{2}\Psi}{\partial\eta^{2}} + \left(1-|\Psi|^{2}\right)\Psi= -k\eta\frac{\partial^{2}\Psi}{\partial s^{2}} + \frac{k}{2}\frac{\partial\Psi}{\partial\eta},
\label{GPE_curv_coordinates_2}
\end{equation}
where $\eta$ represents the distance along the normal direction from the supporting line such that

\begin{equation}
\vec{r} = \vec{r}_{0}(s,t) + \eta\vec{n}_{0}(s,t),
\label{trans_curvilinear_coordinates}
\end{equation}
with $\vec{n}_{0}(s,t)$  the unit vector normal to the curve $\vec{r}_{0}(s,t)$.

Starting from Eq.\ref{GPE_curv_coordinates_2}, we now investigate the modifications introduced by curvature on the solitonic solution along the curved line. As a starting point, we consider the zeroth-order asymptotic approximation ($\Psi \approx \Psi_{0}$), corresponding to the curvature-free case ($k=0$). In this limit, we recover the standard one-dimensional Gross-Pitaevskii equation:

\begin{equation}
    i\frac{\partial\Psi_{0}}{\partial t} + \frac{1}{2}\frac{\partial \Psi_{0}}{\partial s^{2}} + \left(1 - |\Psi_{0}|^{2}\right)\Psi_{0} = 0.
    \label{GPE_curv_coordinates_2_for_Psi_Psi_0}
\end{equation}

Changing Eq.\ref{GPE_curv_coordinates_2_for_Psi_Psi_0} into the soliton co-moving frame by applying a change of variables $\eta = s - vt$, we rewrite the equation as:

\begin{equation}
    -iv\frac{\partial\Psi_{0}}{\partial\eta} + \frac{1}{2}\frac{\partial^{2}\Psi_{0}}{\partial\eta^{2}} + \left(1-|\Psi_{0}|^{2}\right)\Psi_{0} = 0,
    \label{GPE_curv_coordinates_2_for_Psi_Psi_0_eta}
\end{equation}
which admits the following wavefunction as a solution:
\begin{equation}
    \Psi_{0}(\eta) = \sqrt{1-v^{2}}\tanh(\sqrt{1-v^{2}}\eta) + iv,
    \label{Psi_0_eta}
\end{equation}
and precisely characterizes a dark soliton. 

To incorporate curvature effects, we consider a first-order correction in the form $\Psi = \Psi_{0}+\nu\Psi_{1}$. The conditions of existence and uniqueness of such solution are discussed in \ref{app:freholm}.

By applying kinematic arguments and expanding the dynamics to leading order in $\nu$, one obtains a system of coupled equations governing the evolution of the curvature $\kappa$ and soliton velocity $v$ along the arc-length parametrization:

\begin{equation}
    \frac{\partial \kappa}{\partial t} - \left(\int_0^s v\kappa \, d\chi\right)\frac{\partial \kappa}{\partial s} = \kappa^2 v + \frac{\partial^2 v}{\partial s^2}
    \label{eq:curvature},
\end{equation}

\begin{equation}
    \frac{\partial v}{\partial t} - \left(\int_0^s v\kappa \, d\chi\right)\frac{\partial v}{\partial s} = -\kappa\left(\frac{1 - v^2}{3}\right)
    \label{eq:velocity}.
\end{equation}

These results establish the theoretical foundation for exploring how curvature affects the propagation of dark solitons in curved geometries. The perturbative approach outlined above highlights the interplay between the intrinsic parameters of the soliton and the local curvature of the supporting manifold. This framework paves the way for exploring the dynamical behavior of dark solitons under varying curvature profiles, as will be discussed in the following sections. 

\section{Constant curvature: solitons in a ring}

\begin{figure}[htbp]
	\centering
    \includegraphics[width=\textwidth]{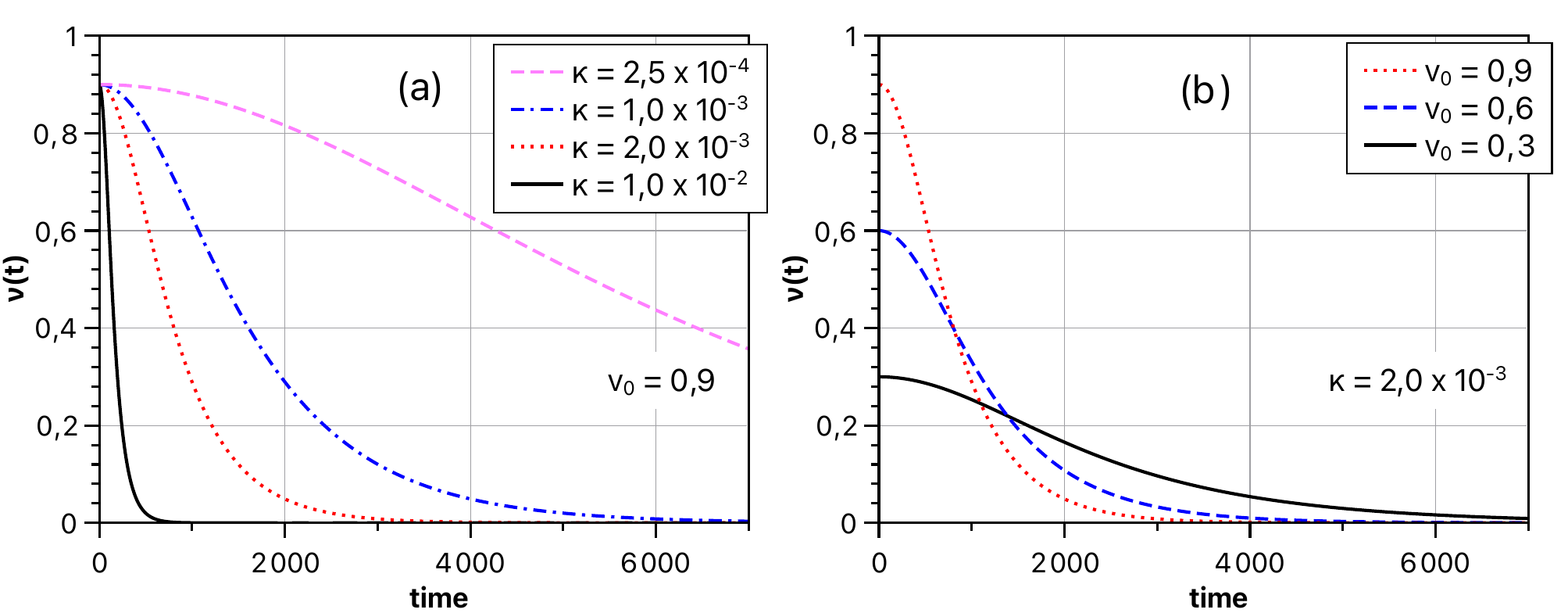}
	\caption{Velocity as a function of time $v(t)$ of a dark soliton with initial velocity $v_0$ along a line with constant curvature $\kappa$. (a) Fixed initial velocity $v_0=0,9$ for several $\kappa$. (b) Constant curvature $\kappa=2,0\times10^{-3}$ for several $v_0$. The red dotted line is the same on both plots, $v_0=0,9$ and $\kappa=2,0\times10^{-3}$.}
	\label{fig:circle}
\end{figure}

In order to understand the non-linear dynamics of the motion of a dark soliton in a BEC with curvature we solve numerically Eqs.\ref{eq:curvature} and \ref{eq:velocity} for a soliton propagating with different initial velocities $v_0$ in a circle of radius R, i.e., a line with constant curvature $\kappa=1/R$. Fig.\ref{fig:circle}a and \ref{fig:circle}b show our results for a given fixed initial velocity and for several curvatures and for a fixed curvature and for several initial velocity, respectively.

At first glance, a few key features stand-out: 
\begin{enumerate}[(i)]
    \item the velocity of the soliton decays from the initial value $v_0$ to zero with a inverted S-shaped profile,
    \item the larger the curvature the faster the decay (see Fig.\ref{fig:circle}a). Indeed, we have explicitly checked that at $\kappa=0$ the velocity of the soliton is kept constant in time as expected from a standard soliton behavior,
    \item the larger the initial velocity $v_0$ the faster the decay 
    (Fig.\ref{fig:circle}b),
    \item the velocity decay rate is larger at shorter times and smaller at the end, given a constant curvature and initial velocity.
\end{enumerate}
A more detailed analysis hints at additional qualitative behavior
\begin{enumerate} [(i)]
\setcounter{enumi}{4}
    \item the fast decay at shorter times is mostly linear and
    \item the final tail of the decay is slower and displays an exponential-like behavior.
\end{enumerate}

To gain physical insight into how curvature affects the soliton velocity dynamics, we adopt a reflected Gompertz function as an \textit{ad hoc} model.
\begin{equation}
    v(t)=a\left\{1-\exp\left[-b\exp(-ct)\right]\right\}
    \label{eq:gompertz}
\end{equation}

This model accounts for the many qualitative features listed above: is an assymetric S-shaped curve with fast and linear decay for small $t$ and exponential decay for large $t$. In specific
\begin{equation}
    \begin{cases} 
      v(t)=a\left[1-\exp(-b)\right]-\left[abc\exp(-b)\right]t & t\xrightarrow{} 0 \\
     v(t)=ab\exp\left(-ct\right) & t\xrightarrow{} \infty \\
   \end{cases}
\label{eq:assymptotic}
\end{equation}

We have adjusted Eq.\ref{eq:gompertz} for soliton dynamics under curvatures ranging from $1,0\times10^{-4}<\kappa<0,2$ and initial velocities from $0,1<v_0<0,9$. As we show below, we find that for a very broad range of constant curvatures the soliton velocity dynamics is fully accounted by the initial velocity and the curvature only, i.e., the dynamics is predictable without free parameters.

We first identify that $c=v_0\kappa$ for any $v_0$ and $\kappa$ in the explored range. So, the long tail slow decay of the soliton velocity (Eq.\ref{eq:assymptotic}) is governed by the curvature and initial velocity only in a very simple functional form.

\begin{figure}[htbp]
	\centering
    \includegraphics[width=\textwidth]{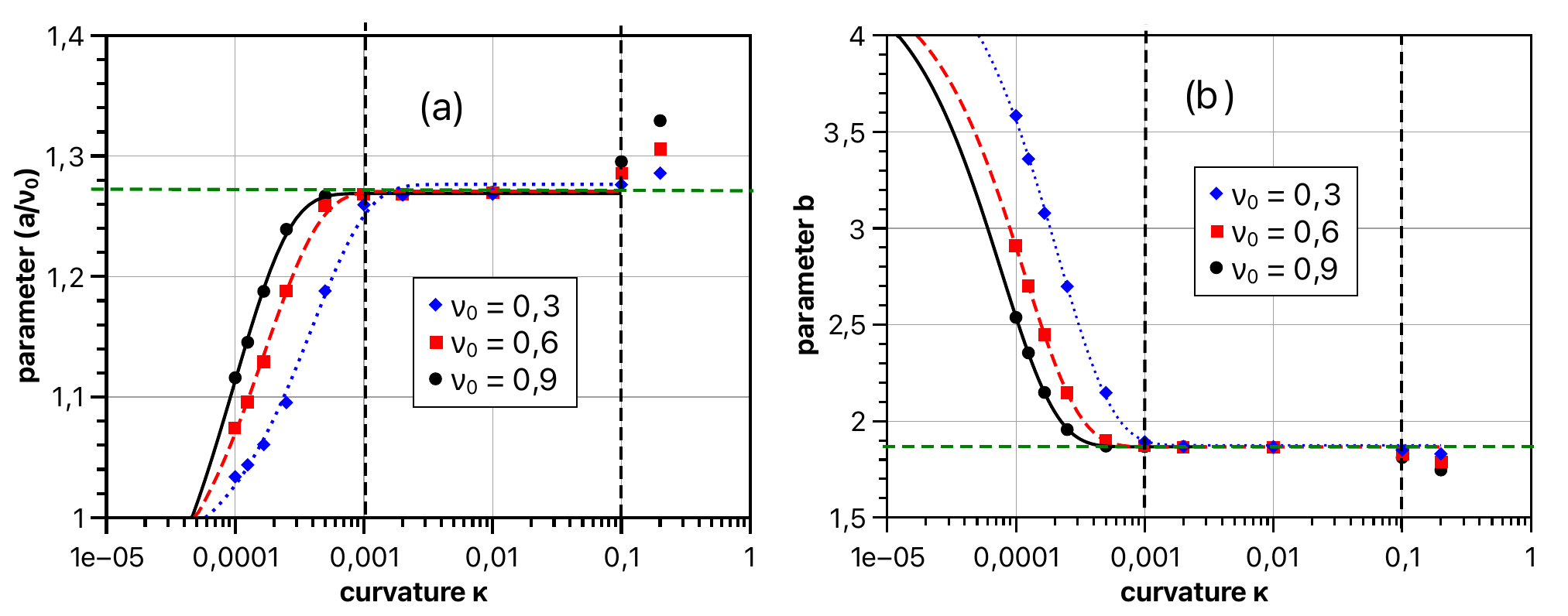}
	\caption{Evolution of parameters $a$ and $b$ of the Gompertz function as a function of the curvature $\kappa$ for several initial velocity $v_0$. (a) Parameter $a/v_{0}$ as a function of $\kappa$. The color (patterned) lines are exponential fittings to the data for $\kappa<0,1$. (b) Parameter $b$ as a function of $\kappa$. The color (patterned) lines are exponential fittings to the data for $\kappa<0,1$. On both: the horizontal green (dashed) line is a guide to the eye to highlight the region where $a/v_{0}$ and $b$ are constant. The range of constant parameters is delimited by the vertical grey (dashed) lines, beteween $\kappa\approx0,001$ and $\kappa\approx0,1$.}
	\label{figandb}
\end{figure}

We then turn our attention to parameters $a$ and $b$. It turns out that, in a range of curvatures that span over almost two orders or magnitude, from $\kappa\approx1,0\times10^{-3}$ to $\kappa\approx1,0\times10^{-1}$ the $b$ parameter is a constant $b=1,87\pm2,0\times10^{-3}$ and, in the same range, the parameter $a$ is directly proportional to $v_0$, as $a=1,27v_0\pm8,0\times 10^{-3}$. Fig.\ref{figandb} displays the behavior in the full range of curvatures.

As it can be seen in Fig.\ref{figandb} for very small ($\kappa<1,0\times10^{-3}$) and very large ($\kappa>0,1$) curvatures, the parameters $a/v_{0}$ and $b$ are not constant. For large curvatures we are approaching the limits where $\nu\xrightarrow{}1$, reflecting the fact that we are breaking the initial assumptions of small curvature. For $\kappa<1,0\times10^{-3}$ the behavior of the parameters $a/v_{0}$  and $b$ deviate from the constant value but are still well explained by a simple exponential behavior as highlighted in Fig.\ref{figandb} by the lines following the data. The behavior in this region is beyond the scope of the present work as it adds an additional layer of non-linear behavior. Therefore, in what follows throughout the present work we restrict ourselves to the region where the parameters are constant.

It is possible then to summarize that the dynamic behavior of the velocity of a dark soliton moving along a line of constant curvature $\kappa$ is fully defined by its initial velocity $v_0$ and the curvature of the line itself for a large range initial velocities and curvatures spanning 2 orders of magnitude. The resulting expression for a constant curvature $\kappa$ is given by

\begin{equation}
    v(t)=1,27v_0\left\{1-\exp\left[-1,87\exp(-\kappa v_0t)\right]\right\}
    \label{eq:constantK}
\end{equation}
with the respective asymptotic limits given by
\begin{equation}
    \begin{cases} 
      v(t)=1,07v_0-\left(0,37v_0^2\kappa\right) t & t\xrightarrow{} 0 \\
     v(t)=2,38v_0\exp\left(-v_0\kappa t\right) & t\xrightarrow{} \infty \\
   \end{cases}
\label{eq:limits}
\end{equation}

\begin{figure}[htbp]
	\centering
    \includegraphics[width=\textwidth]{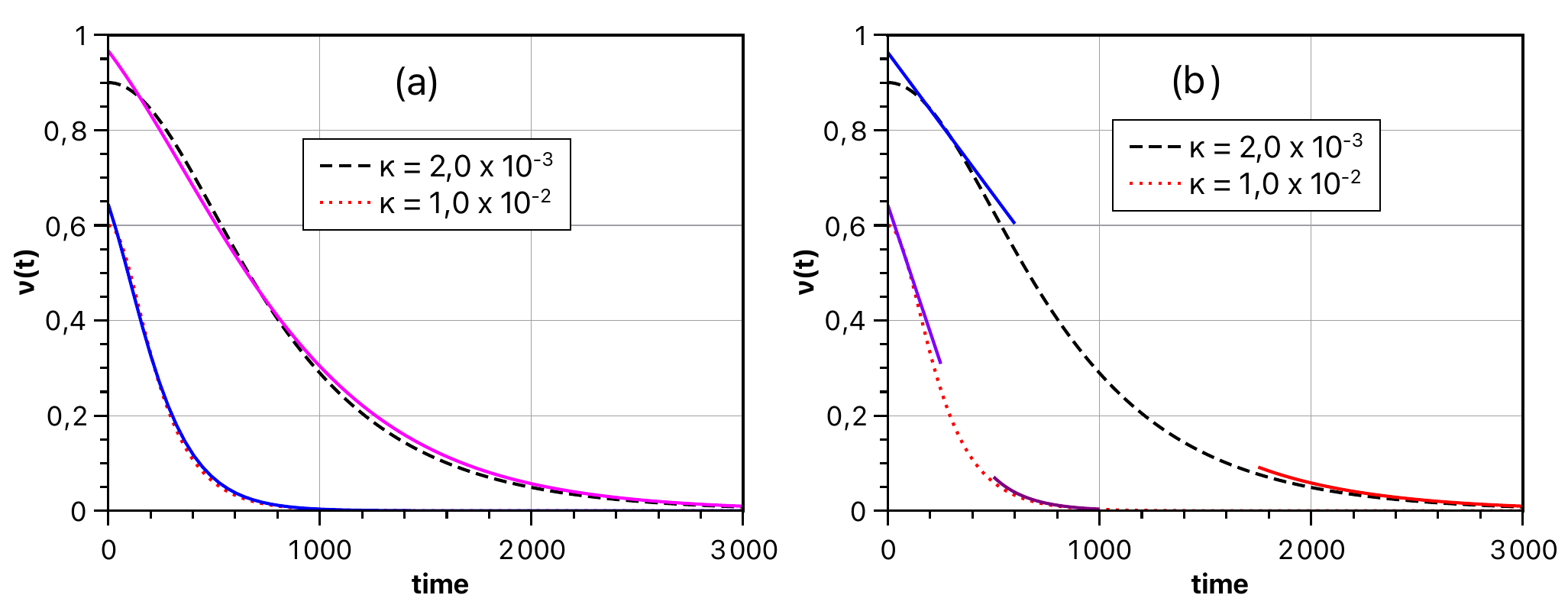}
	\caption{Velocity as a function of time $v(t)$ of a dark soliton with initial velocity $v_0=0,9$ and $v_0=0,6$. (a) Dashed and dotted lines are the numerical solution of Eq.\ref{eq:velocity} and solid lines are the model of Eq.\ref{eq:constantK} showing the very goood agreement. (b) Same as (a) but with the assymptotic behavior of Eq.\ref{eq:limits} for small and large $t$, explicitly highlighting the initial linear decay and the exponential behavior for large $t$.}
	\label{fig:limits}
\end{figure}

The asymptotic limits capture the fast linear initial decay, proportional to the square of the initial velocity and the curvature. Also, for $t=0$ it overshoots the initial velocity value by only $\approx7\%$. The final, slower decay, has a time constant proportional to the product of $v_0$ and $\kappa$. This behavior is illustrated in Fig.\ref{fig:limits}a, with the full model of Eq.\ref{eq:constantK} and in Fig.\ref{fig:limits}b with the asymptotic limits, for two values of curvature $\kappa$ inside the region where Eq.\ref{eq:constantK} and Eq.\ref{eq:limits} are defined. Therefore, eqs.\ref{eq:constantK} and \ref{eq:limits} are the starting point to investigate, in the next section, the behavior of $v(t)$ as the dark soliton moves along a line of varying $\kappa$.

Surprisingly, an additional feature emerges from the behavior described above: the angular displacement of the soliton from its initial position to the point where it comes to rest remains constant, independent on the curvature across approximately two orders of magnitude in curvature and a factor of 10 in initial velocity. In fact, within the range of validity of Eq.\ref{eq:constantK}, this angular displacement under constant curvature behaves as a constant of motion, with a value of $\theta = 1.602$ rad, which is very close to $\pi/2$. This result is supported both by the \textit{ad hoc} model and by direct numerical evaluation of the density distribution derived from Eq.\ref{Psi_0_eta}, tracking the soliton until its velocity becomes significantly smaller than the initial velocity.

\section{Solitons along an ellipse: curvature that changes in space}

We now explore a scenario in which the curvature $\kappa$ is well-defined but varies as the soliton propagates. Specifically, we analyze the behavior of $v(t)$ as the soliton moves along an ellipse, whose curvature is given by

\begin{equation}
\kappa\left(\theta\right) = \frac{\sqrt{1-e^2}}{R\left[\left(1-e^2\right)\cos^2(\theta) + \sin^2(\theta)\right]^{3/2}},
\label{curvature_elipse}
\end{equation}
where $e = \left(1 - \rho^{2}/R^{2}\right)^{1/2}$ is the eccentricity of the ellipse, and $R$ and $\rho$ are its semi-major and semi-minor axes, respectively. The angle $\theta$ denotes the position along the ellipse at which the local curvature is computed; $\theta = 0$ ($\theta = \pi/2$) corresponds to the direction along the major (minor) axis. Throughout this section, we fix $R = 200$, ensuring that for $e \leq 0.9$, the local curvature $\kappa(\theta)$ remains within the range displayed in Fig.~\ref{figandb}, where our model with constant coefficients remains valid across the full $\theta$ range. 

\begin{figure}[htbp]
	\centering
    \includegraphics[width=\textwidth]{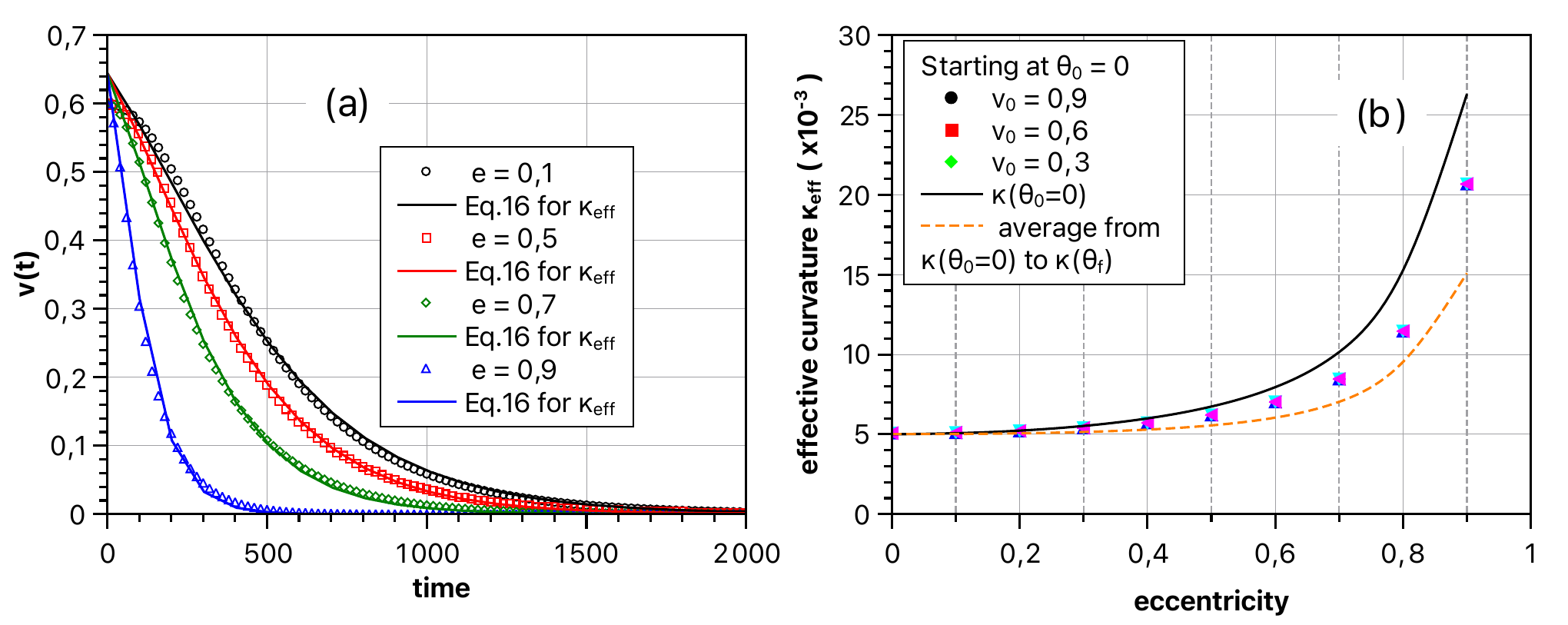}
	\caption{(a) Velocity as a function of time $v(t)$ of a dark soliton with initial velocity $v_0=0,6$ moving along a line with curvature defined by the ellipse curvature (Eq.\ref{curvature_elipse}) for several different eccentricities $e$, starting at $\theta_0=0$ and the corresponding adjust of the model of Eq.\ref{eq:constantK} for an effective curvature $\kappa_{\text{eff}}$. (b) Effective curvature $\kappa_{\text{eff}}$ for several initial velocities $v_0$ as a function of eccentricity, black circle $v_0=0,9$; red square  $v_0=0,6$; green diamond  $v_0=0,3$; extracted from adjust of Eq.\ref{eq:constantK} to data like the displayed in (a). Also shown: black solid line: curvature at $\theta_0=0$; orange dashed line: average curvature from $\kappa(\theta = 0)$ to $\kappa(\theta = \theta_f$) (see text).
    }
	\label{fig:theta00}
\end{figure}

Since $\kappa$ varies with $\theta$, we begin by exploring the behavior of $v(t)$ for different initial positions $\theta_0$ along the ellipse. Starting at $\theta_0 = 0$, \textit{i.e.}, along the major axis, we compute $v(t)$ for several initial velocities $v_0$, spanning a range of eccentricities $0.1 \leq e \leq 0.9$. The behavior resembles that observed in Fig.\ref{fig:circle}: as the eccentricity increases, the initial local curvature increases as well, leading to a faster decay of the soliton velocity. This is illustrated in Fig.\ref{fig:theta00}a, which shows $v(t)$ for $v_0 = 0.6$ and varying eccentricities.

To model this dynamics, we fit the solution of Eq.\ref{eq:constantK} with an effective curvature, $\kappa_{\text{eff}}$, which characterizes the soliton motion despite $\kappa(\theta)$ being not a constant anymore. Remarkably, although the soliton initially experiences a curvature that can vary by a factor of $\approx5$\footnote{$\kappa\left(\theta=0;e=0,1\right)=0,005$ and $\kappa\left(\theta=0;e=0,9\right)=0,026$.} and decreases continuously as it moves, Eq.\ref{eq:constantK} still provides a very good description to the dynamics. Fig.\ref{fig:theta00}b presents the extracted values of $\kappa_{\text{eff}}$ as a function of eccentricity, for several initial velocities $v_0$.

We observe that $\kappa_{\text{eff}}$ approaches $\kappa = 5 \times 10^{-3}$ for small eccentricities, as the ellipse with fixed major semi-axis $R = 200$ increasingly resembles a circle of constant radius. Notably, $\kappa_{\text{eff}}$ is independent of the initial velocities $v_0$. 

To investigate the physical meaning of $\kappa_{\text{eff}}$, we plot alongside $\kappa_{\text{eff}}$ in Fig.\ref{fig:theta00}b the initial curvature $\kappa(\theta = 0)$ as a function of eccentricity, shown as the solid black line, and a simple average of the curvature from $\kappa(\theta = 0)$ to $\kappa(\theta = \theta_f$) (orange line), where $\theta_f$ is the final position at which the soliton comes to rest. We evaluate $\theta_f$ by following the soliton along the elliptical path, computing the density distribution as a function of time using Eq.\ref{Psi_0_eta}, halting the evolution when $v(t) \leq 1 \times 10^{-4}$, more than three orders of magnitude smaller than any initial velocity $v_0$ considered. It becomes immediately apparent that $\kappa_{\text{eff}}$ is not matched by any of the methods. The fact that neither simple procedures of estimating $\kappa_{\text{eff}}$ capture its value is an indication that, although the dynamics is clearly defined by the local curvature and one can assign an effective value for it, the exact dependence is due to a more complex dynamics. Nevertheless, similarly to $\kappa_{\text{eff}}$, the final position of the soliton $\theta_f$ is also independent of the initial velocity.

\begin{figure}[htbp]
	\centering
    \includegraphics[width=\textwidth]{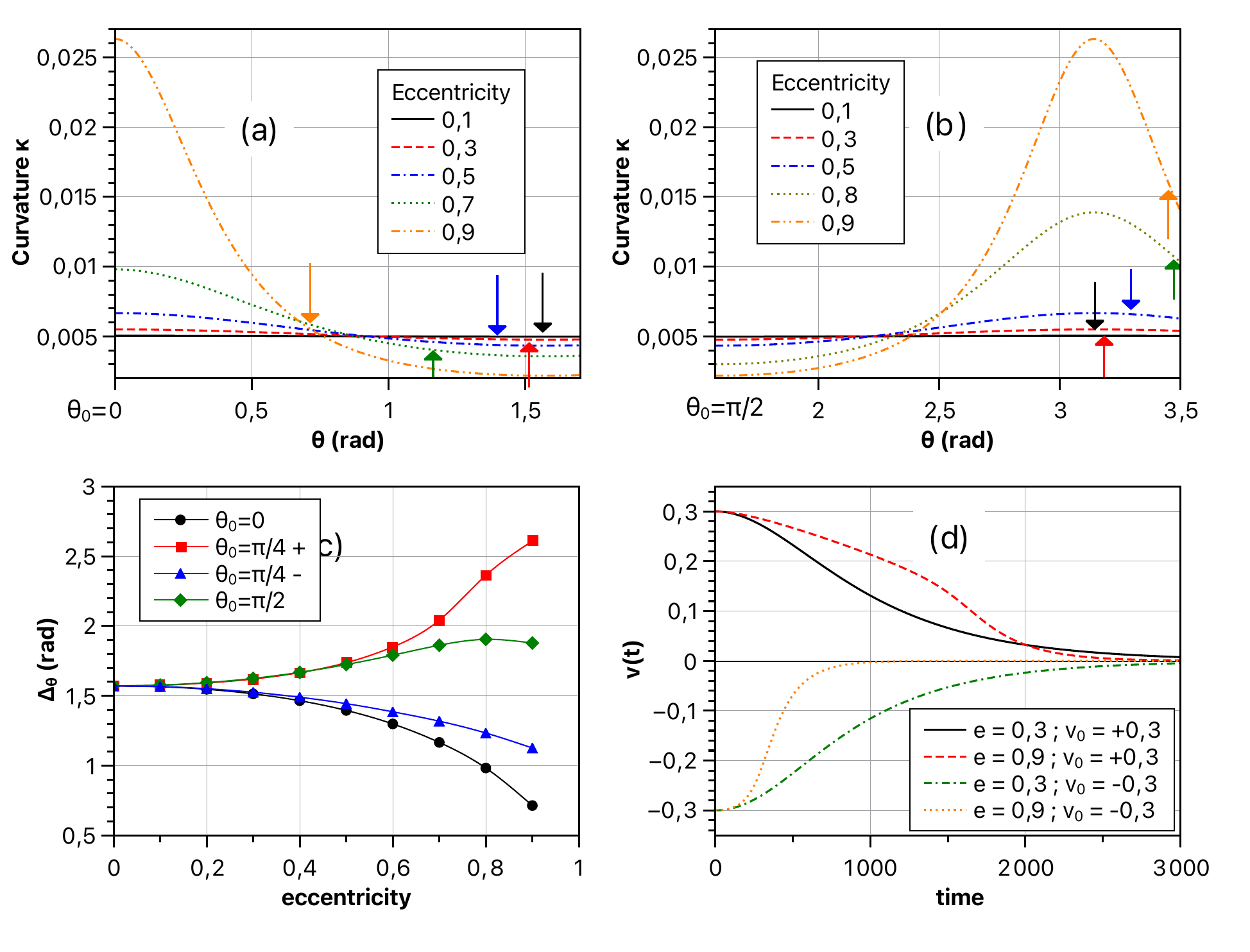}
	\caption{(a) Curvature profiles of elliptical trajectories for various eccentricities. A soliton launched from $\theta_0 = 0$ with any given initial velocity comes to rest at the positions indicated by arrows. The arrow colors correspond to the eccentricity curves shown. (b) Same as (a), but for solitons launched from $\theta_0 = \pi/2$. (c) Angular displacement $\Delta_\theta$ of the soliton, measured from its initial to final position along elliptical trajectories, as a function of eccentricity for different initial angles $\theta_0$. Black circles: $\theta_0 = 0$; green diamonds: $\theta_0 = \pi/2$; red squares: $\theta_0 = \pi/4$ moving towards increasing $\theta$ (i.e., decreasing curvature); blue triangles: $\theta_0 = \pi/4$ moving towards decreasing $\theta$ (i.e., increasing curvature). In all cases, the angular displacements converge toward the constant-curvature value (see text). (d) Representative velocity profiles $v(t)$ of a dark soliton launched from $\theta_0 = \pi/4$ in two opposite directions. The distinct behavior illustrates the asymmetric influence of the curvature landscape on soliton dynamics.}
	\label{fig:thetaf}
\end{figure}

We now compute the soliton dynamics for $\theta_0 = \pi/2$, \textit{i.e.}, when the soliton starts at a point along the minor axis of the ellipse. In this case, the behavior becomes less straightforward as the eccentricity increases. Near $\theta = \pi/2$, the dependence of the initial curvature $\kappa$ on the eccentricity $e$ is reversed: as $e$ increases, the initial curvature decreases (see Fig.\ref{fig:thetaf}b). However, the variation in the initial curvature at $\theta = \pi/2$ spans only a factor of about $\approx2.5$\footnote{$\kappa\left(\theta=\pi/2;e=0,1\right)=0,005$ and $\kappa\left(\theta=\pi/2;e=0,9\right)=0,002$.}. This difference diminishes as the soliton moves away from its initial position, inverts with the larger eccentricity taking over the larger local curvature $\kappa$ and then quickly increases towards its maximum value around $\theta=\pi$. Indeed, the soliton dynamics reflects this behavior in the curvature.

\begin{figure}[htbp]
	\centering
    \includegraphics[width=\textwidth]{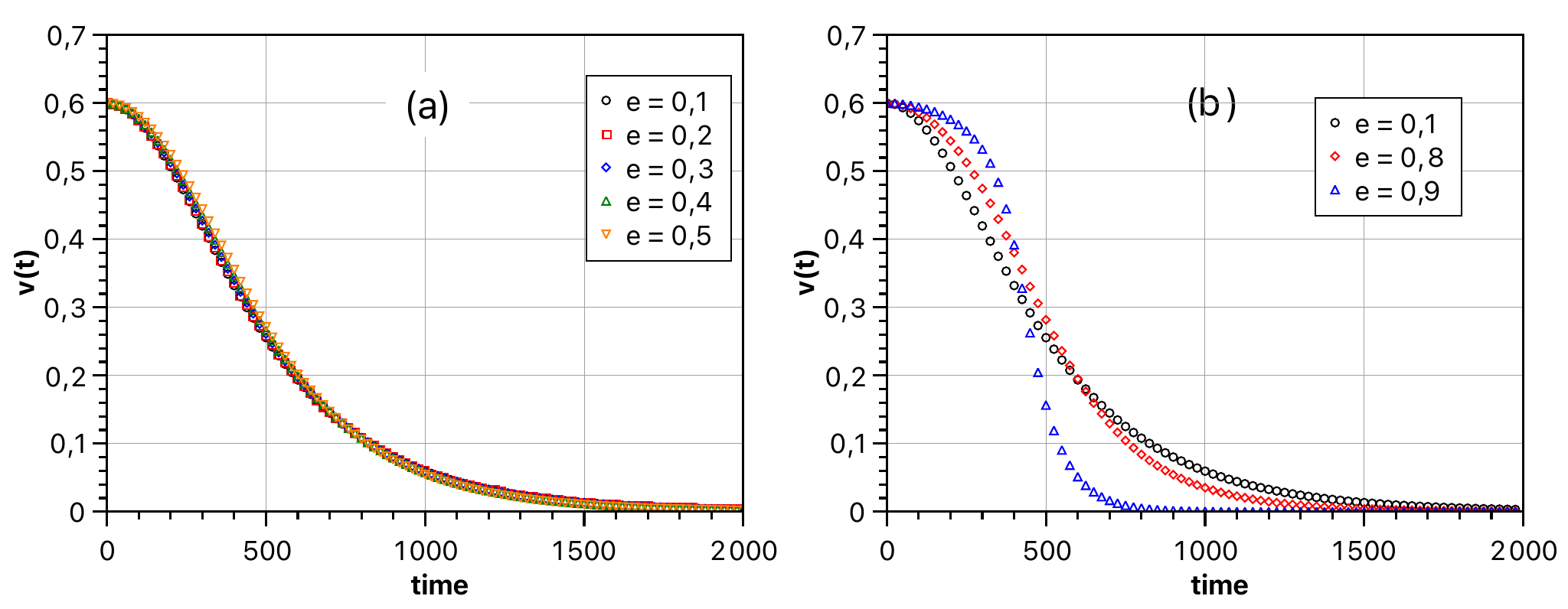}
	\caption{(a) velocity as a function of time $v(t)$ of a dark soliton with initial velocity $v_0=0,6$ moving along a line with curvature defined by the ellipse curvature Eq.\ref{curvature_elipse} for several small eccentricities $e$, starting at $\theta_0=\pi/2$ and the corresponding adjust of the model of Eq.\ref{eq:constantK} for an effective curvature $\kappa_{\text{eff}}$. (b) Same as (a) comparing small ($e=0,1$) and large ($e=0,8$ and $e=0,9$) eccentricities, showing the change in behavior of the latter due to the rapid change in local curvature (see text).}
	\label{fig:thetapi2}
\end{figure}

Figure \ref{fig:thetapi2}a displays $v(t)$ as a function of time for small eccentricities ($e \leq 0.5$), where the local curvature remains small and approximately constant over the trajectory segment. In this regime, the soliton dynamics are effectively captured by Eq.\ref{eq:constantK}, with extracted effective curvature values lying within a narrow interval,  $4.93 \times 10^{-3} \leq \kappa_{\text{eff}} \leq 5.07 \times 10^{-3} $, in close agreement with the local geometric curvature $\kappa \approx 5 \times 10^{-3}$.

At higher eccentricities, the curvature at $\theta = \pi/2$ is smaller, as previously discussed, resulting in an initially slower decay of $v(t)$, consistent with the global behavior presented in Fig.\ref{fig:theta00}. This trend, however, inverts at later times: for large $e$, the soliton velocity exhibits a rapid decay phase, as shown in Fig.\ref{fig:thetapi2}b, which contrasts the time evolution for small and large eccentricities. The dynamics thus reveal a two-stage structure: an initial regime dominated by low curvature and gradual decay, followed by an abrupt exponential drop in velocity. This transition is not accounted for within the framework of the single-parameter model, which assumes constant curvature, even if effective. Nevertheless, the overall behavior of $v(t)$ remains closely linked to the local curvature profile, suggesting that the soliton dissipative dynamics are still predominantly governed by geometric effects even in regimes where curvature varies significantly.

To explicitly illustrate the curvature landscape experienced by the soliton and its influence on the velocity decay, we track its trajectory from the initial position $\theta_0 = 0$ ($\theta_0 = \pi/2$) to the final rest position $\theta_f$, as indicated in Fig.\ref{fig:thetaf}a (b) for various eccentricities. For $\theta_0 = 0$, the total angular displacement decreases with curvature: the higher the overall curvature along the path, the less the soliton travels before coming to rest. Besides, the variation of curvature along the soliton trajectory is predominantly monotonic, decreasing for most of the trajectory. The final position $\theta_f$ is indicated by an arrow-matching color in Fig.\ref{fig:thetaf}a.

In contrast, for $\theta_0 = \pi/2$, shown in Fig.\ref{fig:thetaf}b, the curvature landscape is nearly flat for small eccentricities, resulting in uniform dynamics. However, for large eccentricities, the soliton begins in a region of minimal curvature, then experiences increasing curvature up to a maximum, followed by a decrease near the end of the trajectory. Notably, for $e = 0,9$, the curvature encountered becomes sufficiently large that the soliton comes to rest after a smaller angular displacement $\Delta_\theta = \theta_f - \theta_0$, as compared, for example, to the case of $e = 0,8$, as indicated by the respective arrows. For all evaluated velocities and eccentricities, although the stopping point $\theta_f$ depends on the eccentricity, it remains independent of the initial velocity, consistent with the behavior observed under constant curvature.

Figure \ref{fig:thetaf}c summarizes the angular displacement $\Delta_\theta$ as a function of eccentricity for several initial positions $\theta_0$. In particular, we also present results for $\theta_0 = \pi/4$ with both positive and negative initial velocities, to highlight the asymmetry in the curvature landscape on either side of this point. The soliton angular displacement clearly differs depending on the direction of motion, reflecting the spatial variation of the local curvature. For small $e$, all trajectories converge to $\Delta_\theta \approx 1,6$ rad, regardless of $\theta_0$, mirroring the motion at constant $\kappa$.

Figure \ref{fig:thetaf}(d) shows representative velocity profiles $v(t)$ for solitons launched from $\theta_0 = \pi/4$, illustrating the distinct dynamical behaviors that arise due to curvature asymmetry. For the case of $e = 0,3$, the constant-curvature model of Eq.\ref{eq:constantK} accurately reproduces the soliton dynamics for both positive and negative initial velocities, despite their qualitative differences, which reflect the asymmetry of the underlying curvature landscape. We have explicitly verified that the constant $\kappa_{\text{eff}}$ model remains valid for $e \leq 0.7$ for positive initial velocities and for $e \leq 0.5$ for negative ones. These ranges correspond to cases where the curvature is either monotonically increasing (negative velocities) or remains small and slowly varying, though not strictly monotonic (positive velocities).

For larger eccentricities the curvature varies more rapidly along the trajectory and the soliton typically crosses the point of maximum curvature (see Fig.\ref{fig:thetaf}c). This leads to a more complex curvature landscape, which in turn induces dynamics that deviate significantly from those predicted by the single-parameter $\kappa_{\text{eff}}$ model. In particular, the curve for $e = 0.9$ with negative initial velocity\footnote{In Fig.\ref{fig:thetaf}d, we show the case $e = 0.9$ and $v_0 = -0.3$, but the qualitative behavior remains the same for any initial velocity $|v_0| \leq 0.9$.} exhibits a markedly nonuniform decay, indicative of multiple dynamical regimes emerging along the soliton trajectory.

\section{Conclusions}

We have investigated the dynamics of a dark soliton propagating in a one-dimensional Bose-Einstein condensate subject to a smooth spatial curvature. Our analysis reveals that the soliton motion is non-trivial yet entirely governed by the local curvature and the soliton initial velocity.

In the case of constant curvature, we find that the soliton velocity initially decays linearly in time, with a rate proportional to the curvature and the square of the initial velocity. At later times, the decay becomes exponential, with a characteristic time scale given by the inverse of the product of the curvature and initial velocity. This model accurately describes the dynamics over two orders of magnitude in curvature and a broad range of initial velocities. Remarkably, under constant curvature, the soliton trajectory spans a fixed angular displacement, amounting to approximately a quarter of a circle, regardless of its initial speed.

We have further extended this framework to spatially varying curvature, specifically considering curvature profiles corresponding to elliptical trajectories. In regimes where the curvature is small, slowly varying, or monotonic, the dynamics remain well-described by an effective constant curvature. In more complex scenarios where a single effective parameter does not suffice, the time-dependent soliton velocity still reflects the influence of the local curvature. Notably, in all cases studied, the total angular distance traveled by the soliton is independent of the initial velocity, emerging as a robust geometric invariant of the motion.

These results establish a quantitative foundation for understanding the influence of geometry on soliton dynamics. They open new paths for the study of nonlinear excitations in curved quantum systems and serve as a starting point for exploring soliton motion in higher-dimensional geometries with spatial curvature.\medbreak

\noindent{\bf CRediT authorship contribution statement} \medbreak

\noindent \textbf{Emanuel A. L. Henn:} Conceptualization, Writing, Data Analysis.\\ \textbf{Jorge A. G. Attie:} Investigation, Methodology, Data Analysis, Writing.\medbreak

\noindent{\bf Declaration of competing interests}\medbreak

The authors declare that they have no known competing financial interests or personal relationships that could have appeared to influence the work reported in this paper.\medbreak

\noindent{\bf Data availability}\medbreak

Numerical data generated for the research described in the article is available on a reasonable request.\medbreak

\noindent{\bf Acknowledgments}\medbreak

This work has received support from São Paulo state funding agency FAPESP Grant 2013/07276-1. This work was carried out with the support of the Coordination for the Improvement of Higher Education Personnel – Brazil (CAPES) – Financing Code 001. \medbreak

\appendix

\section{Conversion of the Gross-Pitaevskii equation to its adimensional representation} 
\label{app:admi}

We start from the Gross-Pitaevskii equation in its dimensional form: 

\begin{equation}
i\hbar\frac{\partial\Psi}{\partial t} = -\frac{\hbar^{2}}{2m}\nabla^{2}\Psi + g|\Psi|^{2}\Psi
\end{equation}
where $m$ is the atomic mass and $g$ is the effective interaction strength. To express it in its adimensional representation we proceed as follows.

First we introduce the following dimensionless variables:

\begin{equation}
\vec{r}' = \frac{\vec{r}}{\xi}, \quad t' = \frac{t}{\tau},
\label{dimensionless_transformation_parameters}
\end{equation}
where $\xi$ and $\tau$ are characteristic length and time scales. Under these transformations, the derivatives transform as:

\begin{equation}
    \frac{\partial}{\partial t} = \frac{1}{\tau}\frac{\partial}{\partial t'}\text{,  } \nabla^{2} = \frac{1}{\xi^{2}}\nabla^{\prime 2}.
    \label{derivatives_transformed}
\end{equation}
Substituting these expression into the Eq. \ref{GPE_dimensional} yields:
\begin{equation}
    i\hbar\frac{1}{\tau}\frac{\partial\Psi}{\partial t'} = -\frac{\hbar^{2}}{2m}\frac{1}{\xi^{2}}\nabla^{\prime 2}\Psi + g|\Psi|^{2}\Psi.
    \label{GPE_dimensionless_intermediate}
\end{equation}
Additionally, the wavefunction is rescaled as:
\begin{equation}
    \Psi' = \frac{\Psi}{\sqrt{n_{0}}},
    \label{wavefunction_rescaled}
\end{equation}
where $n_{0}$ is the background density of the condensate. Equation \ref{GPE_dimensionless_intermediate} then becomes:
\begin{equation}
    i\frac{\partial\Psi^{\prime}}{\partial t^{\prime}} = - \frac{\tau\hbar}{2m\xi^{2}}\nabla^{\prime 2}\Psi' + \frac{gn_{0}\tau}{\hbar}|\Psi^{\prime}|^{2}\Psi^{\prime}.
    \label{GPE_dimensionless_intermediate_2}
\end{equation}
In order to render the equation dimensionless, both prefactors must be set to unity, which leads to the following conditions for the characteristic time and length scales:

\begin{equation}
    \tau = \frac{\hbar}{gn_{0}}, \ \xi = \frac{\hbar}{\sqrt{2mgn_{0}}}.
    \label{explicity_form_transf_parameters}
\end{equation}

It is important to emphasize that the scaling factor $\xi$ was initially introduced as an arbitrary characteristic length, without any imposed physical meaning. Interestingly, the dimensional analysis naturally yields an expression for $\xi$ that exactly matches the healing length of the condensate - the length scale over which perturbations in the density recover. Therefore, the appearance of the healing length arises spontaneously from the adimensionalization procedure, rather than being postulated from the outset. 

With these choices, the Gross-Pitaevskii equation take the dimensionless form:


\begin{equation}
    i\frac{\partial\Psi}{\partial t} + \frac{1}{2}\nabla^{2}\Psi + |\Psi|^{2}\Psi = 0.
\end{equation}
where we have already dropped the primes for simplicity. This is the equation shown in the main text.

\section{Gross-Pitaevskii equation in curvilinear coordinates}
\label{app:GPEcurve}

The relation established by Eq. \ref{curvature_polar_coordinates} allows us to derive a connection between the curvature at two nearby points along the curve, namely $k(s_{a},t)$ and $k(s_{a'},t+dt)$, where $ds = s_{a'} - s_{a}$ denotes the infinitesimal displacement of the dark soliton along the curve during a small time interval $dt$.

The radial coordinate at point $a'$ can be approximated by:

\begin{equation}
    \rho_{a'} = \rho(\phi_{a'},t+dt) = \rho(\phi_{a},t) - v(s_{a},t)dt,
    \label{rho_displaced_point}
\end{equation}
where $\rho(\phi_{a},t) = R(s_{a},t) = 1/k(s_{a},t)$ corresponds to the local radius of curvature of the soliton at position $s_{a}$ and time $t$. After performing the required derivatives and extensive algebraic manipulations, the curvature at the displaced point $(s_{a'},t+dt)$ can be expressed as:

\begin{equation}
    k(s_{a'}, t+dt) = \frac{\rho(\phi_{a},t) - 2v(s_{a},t)dt + \partial^{2}_{ss}v(s_{a},t)\rho^{2}(\phi_{a},t)dt}{\rho^{2}(\phi_{a},t) - 3\rho(\phi_{a},t)v(s_{a},t)dt}.
    \label{k_a_prime_equation}
\end{equation}
Since the change in curvature $dk$ can be written both as $dk = k(s_{a'},t+dt) - k(s_{a},t)$, and, from the absolute derivative, $dk = \partial_{s}kds + \partial_{t}kdt$, it follows that:

\begin{equation}
    ds = -\left(\int_{0}^{s}v(s',t)k(s',t)ds'\right)dt,
    \label{ds_expression}
\end{equation}
which relates the infinitesimal displacement $ds$ of the soliton along the curve to the local curvature and velocity.

The analysis of the dynamics of a curved dark soliton requires considering temporal evolution of the curvature and its coupling to the soliton velocity. An infinitesimal displacement along the supporting curve induces a variation in the curvature experienced by the soliton, ultimately leading to a nonlinear transport equation for $k(s,t)$.

\begin{equation}
\frac{\partial k}{\partial t} - \left(\int_{0}^{s}v(s',t)k(s',t)ds'\right)\frac{\partial k}{\partial s} = k^{2}v + \frac{\partial^{2}v}{\partial s^{2}}.
\label{model_equation_1}
\end{equation}

A more appropriate formulation to describe the behavior of these excitations is obtained by adopting curvilinear coordinates $(s,\eta)$. The transition to this coordinate system is expressed as:

\begin{equation}
\vec{r} = \vec{r}_{0}(s,t) + \eta\vec{n}_{0}(s,t),
\end{equation}
where $\vec{n}_{0}(s,t)$ is the unit vector normal to the curve $\vec{r}_{0}(s,t)$, and $\eta$ represents the distance along the normal direction from the supporting line. The Lamé coefficients for this coordinate system are given by:

\begin{equation}
h_{s} = 1 - k\eta \quad \text{and} \quad h_{\eta} = 1.
\end{equation}

In these new coordinates, Eq. \ref{DS_wavefunction_0} becomes:

\begin{equation}
\Psi_{0} = \sqrt{1 - v^{2}}\tanh\left(\sqrt{1 - v^{2}}\eta\right) + iv,
\label{Psi_0_DS_curv_coordinates}
\end{equation}
and serves as a starting point for an asymptotic expansion of the solution:
\begin{equation}
\Psi = \Psi_{0} + \nu\Psi_{1} + \mathcal{O}[\text{sup}],
\label{func_onda_assint}
\end{equation}
where $\nu$ is a small parameter that quantifies the effects of curvature. By rewriting the GPE in $(s,\eta)$ coordinates and the corresponding Lamé coefficients, one can isolate the dominant first-order terms in $\nu$, revealing how the curvature modifies the soliton dynamics. This approach enables the description of not only the soliton propagation, but also the deformations induced by the geometry.

Thus, the GPE can be rewritten in curvilinear coordinates as follows \cite{PhysRevA.85.053620}:

\begin{equation}
i\frac{\partial\Psi}{\partial t} + \frac{1}{2h_{s}h_{\eta}}\left[\frac{\partial}{\partial s}\left(\frac{h_{\eta}}{h_{s}}\frac{\partial\Psi}{\partial s}\right)+\frac{\partial}{\partial\eta}\left(\frac{h_{s}}{h_{\eta}}\frac{\partial\Psi}{\partial\eta}\right)\right] + \left(1 - |\Psi|^{2}\right)\Psi = 0,
\label{GPE_curv_coordinates}
\end{equation}
substituting the Lamé coefficients yields:

\begin{equation}
i\frac{\partial \Psi}{\partial t} + \frac{1}{2}\frac{\partial^{2}\Psi}{\partial s^{2}} + \frac{k}{2}\frac{\partial^{2}\Psi}{\partial\eta^{2}} + \left(1-|\Psi|^{2}\right)\Psi= -k\eta\frac{\partial^{2}\Psi}{\partial s^{2}} + \frac{k}{2}\frac{\partial\Psi}{\partial\eta}.
\end{equation}

\section{Existence and uniqueness of the first order correction of the soliton solution} \label{app:freholm}

We start with the GPE in cuvulinear coordinates
\begin{equation}
    iv\frac{\partial\Psi_{0}}{\partial\eta} + \frac{1}{2}\frac{\partial^{2}\Psi_{0}}{\partial\eta^{2}} + \left(1-|\Psi_{0}|^{2}\right)\Psi_{0} = 0,
\end{equation}
the corresponding dark soliton solution

\begin{equation}
    \Psi_{0}(\eta) = \sqrt{1-v^{2}}\tanh(\sqrt{1-v^{2}}\eta) + iv,
\end{equation}
and first-order correction in the form $\Psi = \Psi_{0}+\nu\Psi_{1}$. 

Substituting this expansion into Eq.\ref{GPE_curv_coordinates_2}, and after performing the required algebraic simplification 
 we obtain:
\begin{dmath}
    iv\frac{\partial\Psi_{1}}{\partial\eta} + \frac{1}{2}\frac{\partial^{2}\Psi_{1}}{\partial\eta^{2}} + \left(1-2|\Psi_{0}|^{2}\right)\Psi_{1} - \Psi_{0}^{2}\Psi_{1}^{*} = -i\frac{\partial\Psi_{0}}{\partial t} + \frac{k}{2}\frac{\partial\Psi_{0}}{\partial\eta} + i\left(\int_{0}^{s}v(s',t)k(s',t)ds'\right)\frac{\partial\Psi_{0}}{\partial s}.
    \label{GPE_curv_coordiantes_2_for_Psi_Psi_1}
\end{dmath}

To verify the consistence conditions of this approximation, we differentiate the zeroth-order equation (Eq.\ref{GPE_curv_coordinates_2_for_Psi_Psi_0_eta}) with respect to $\eta$, yielding:

\begin{equation}
    -iv\frac{\partial^{2}\Psi_{0}}{\partial\eta^{2}} + \frac{1}{2}\frac{\partial^{3}\Psi_{0}}{\partial\eta^{3}} + \left(1 - 2|\Psi_{0}|^{2}\right)\frac{\partial\Psi_{0}}{\partial\eta} - \Psi_{0}^{2}\frac{\partial\Psi_{0}^{*}}{\partial\eta} = 0,
\end{equation}
which allows us to identify that the correction equation is satisfied if:

\begin{multline}
    \Psi_{1} = \frac{\partial\Psi_{0}}{\partial\eta} \ \text{and} \\ F(\Psi_{0},v,k) :=-\frac{\partial\Psi_{0}}{\partial t} + \frac{k}{2}\frac{\partial\Psi_{0}}{\partial\eta} + i\left(\int_{0}^{s}v(s',t)k(s',t)ds'\right)\frac{\partial\Psi_{0}}{\partial s} = 0.
    \label{uniquiness_existence_conditions}
\end{multline}

Multiplying this last equation by $\Psi_{1}^{*} = \partial\Psi_{0}^{*}/\partial\eta$, integrating the result with respect to $\eta$ and taking the real part of it, we have:
\begin{equation}
   \operatorname{Re}\left\{\int_{-\infty}^{\infty}F(\Psi_{0},v,k)\frac{\partial\Psi_{0}^{*}}{\partial\eta}d\eta\right\} = 0,
\end{equation}

Which corresponds to the Fredholm condition \cite{PhysRevA.85.053620, Kress1999} for existence and uniqueness of a solution applied to the first-order correction of $\Psi$. In particular, the derived condition for solution existence ensures the presence and uniqueness of the first-order correction to the solitonic wavefunction, thus confirming the consistency of the asymptotic expansion.


\end{document}